\documentclass[fleqn,usenatbib]{mnras}
\usepackage{newtxtext,newtxmath}
\usepackage[T1]{fontenc}
\DeclareRobustCommand{\VAN}[3]{#2}
\let\VANthebibliography\thebibliography
\def\thebibliography{\DeclareRobustCommand{\VAN}[3]{##3}\VANthebibliography}
\usepackage[final]{graphicx}	
\usepackage{mwe} 
\usepackage{multirow}
\usepackage{booktabs}
\usepackage[flushleft]{threeparttable}
\usepackage{caption}
\usepackage[flushleft]{threeparttable}
\usepackage{subcaption}
\usepackage{comment}
\usepackage{csquotes}
\usepackage{scrextend}
\usepackage{float}
\usepackage{txfonts}
\usepackage{romannum}
\usepackage{amsmath}
\usepackage{amssymb}
\usepackage{chemmacros}
\usepackage{chemfig}
\title[3-Pyrroline in Interstellar Environments]{Theoretical Investigation of Interstellar 3-Pyrroline: Formation, Rotational and Vibrational Spectroscopy}

\author[Anshika Pandey]{
Anshika Pandey,\thanks{E-mail: 3anshika.1@gmail.com}
Akant Vats, Satyam Srivastav, Amit Pathak\thanks{E-mail: amitpah@gmail.com} and 
K. A. P. Singh
\\
Department of Physics, Institute Of Science, Banaras Hindu University, Varanasi-221005, India\\
}

\date{Accepted XXX. Received YYY; in original form ZZZ}

\pubyear{2023}
\DeclareUnicodeCharacter{2212}{-}
\begin{document}
\label{firstpage}
\pagerange{\pageref{firstpage}--\pageref{lastpage}}
\maketitle

\begin{abstract}
The recent detection of CN-functionalized aromatics partly addresses the long-standing mystery of the apparent absence of five- and six-membered rings in interstellar environments. N-heterocycles, which are crucial as the fundamental structures of nucleobases, have been a focus of these aromatic searches due to their biological significance. Although N-heterocycles have not been conclusively detected in astrophysical environments, their presence in chondrites and meteorites signifies their interstellar and circumstellar connection. Precise spectral data identifies the unique signatures of molecules, confirming their presence in space. 
In this light, the present work reports an extensive computational investigation on interstellar 3-pyrroline; a five-membered ring N-heterocycle. This includes an alternative formation route in cold interstellar environments and highly accurate rotational and vibrational spectroscopy. The results indicate that 3-pyrroline can form on dust grain surfaces from vinyl cyanide, as its formation from pyrrole through double hydrogenation may lead to the formation of pyrrole itself via a H$_{2}$ abstraction process. 3-pyrroline's rotational transition at 52.3 GHz offers a potential tool for its detection in cold interstellar regions. Additionally, the strongest infrared features of 3-pyrroline at 16.09 $\mu$m and $\sim$3.50 $\mu$m are observable with JWST. The provided data is crucial for laboratory identification and future interstellar observations of 3-pyrroline at both radio and IR wavelengths.

\end{abstract}




\section{Introduction}
While different classes and types of organic molecules have been detected in the interstellar medium (ISM) \citep{McGuire_2022}, there was no astronomical signature of branched chain molecules (BCMs) and polycyclic aromatic hydrocarbons (PAHs) until the last decade. Thanks to recent strides in both ground- and space-based radio and infrared telescopes, such as ALMA \citep{2009IEEEP..97.1463W}, GBT \citep{1996AAS...188.4602L}, and JWST \citep{2006SSRv..123..485G}, it is now possible to delve into the evolved realms of interstellar physics and chemistry.
BCM was first detected in the form of \textit{iso}-propyl cyanide (\textit{i}-C$_{3}$H$_{7}$CN) \citep{2014Bell,pagani17}, and recently PAHs have been identified mainly in the form of CN-functionalized aromatics \citep{2018McGuire,2021Sci...371.1265M,2021Cerni,2021ApJ...913L..18B,2021Mcca,2022ApJ...938L..12S}. The detection of both the BCM and CN-PAHs is of great importance because they could potentially serve as a building block for life due to the side chain structure, aromaticity, and the presence of carbon and nitrogen atoms---basic chemical ingredients necessary for life. Having successfully identified several aromatic compounds, the exploration of nitrogen (N)-heterocyclic molecules (where an N atom is a member of a ring) emerges as the next natural progression in unraveling the chemical complexity of interstellar environments \citep[e.g.][]{2022JPCA..126.2716B}.

N-heterocycles hold immense biological importance as they serve as the fundamental structure of nucleobases, which are the building blocks of DNA and RNA responsible for carrying the genetic information of living organisms \citep[e.g.][]{2022JPCA..126.2716B}. However, until now, no five- or six-membered ring heterocycles have been unequivocally identified in the ISM. Several N-heterocycles, for e.g., imidazole, pyridine, pyrimidine, pyrrole, indole, quinoline, and isoquinoline have been searched towards different astronomical regions, with unsuccessful detection \citep[][and references therein]{2022JPCA..126.2716B}. Despite their non-detection, N-heterocycles have been detected in several chondrites \citep{hayatsu1964,folsome1971,folsome1973,hayatsu1975,vander1977,komiya1993}, which strongly signifies their interstellar and circumstellar connection \citep{2022JPCA..126.2716B}. Moreover, observations of PAH IR emission bands strongly support the significant abundance of N-heterocycles in the ISM \citep{2008Tiel}, with the 6.2 $\mu$m PAH emission band in interstellar environments believed to originate from PAH carriers containing nitrogen in their structures. \citep{2005ApJ...632..316H,2021ApJ...923..202R, 2022PASJ...74..161V}.

\begin{figure}
\centering
    \includegraphics[width=0.4\textwidth]{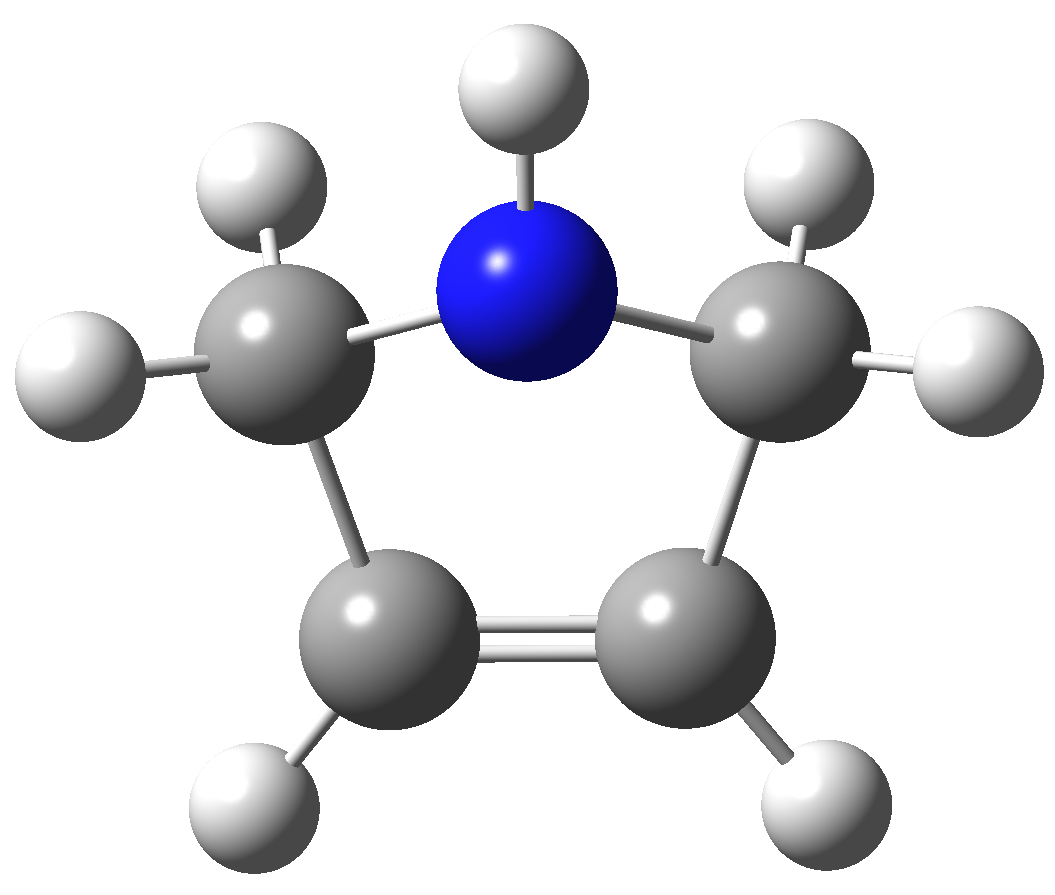}
    \caption{Structural depiction of 3-Pyrroline.}
    \label{fig:1}
\end{figure}
Pyrrole (C$_{4}$H$_{5}$N), a significant interstellar N-heterocycle, has been searched towards the hot core Sgr B2(N) with a column density of  3--6 $\times$ 10$^{13}$ cm$^{-2}$ \citep{1980ApJ...241..155M}, and the cold molecular cloud TMC-1, where the column density is 4 $\times$ 10$^{12}$ cm$^{-2}$ \citep{1980ApJ...242..541K}. Pyrrole can produce several other N-heterocycles in the ISM, such as pyridine (C$_{5}$H$_{5}$N) with methylidyne radical (CH) \citep{2022JPCA..126.2716B} and 2- and 3-pyrrolines (C$_{4}$H$_{7}$N) with its hydrogenation \citep{2021MNRAS.505.3157M}. It is important to note here that both pyrrole and pyrrolines have an NH bond in their structures.

The detection of cyanocyclopentadiene (c-C$_{5}$H$_{5}$CN), a five-membered ring molecule containing a CN group, raises questions about why pyrrole hasn't been observed even though both are found to have comparable column densities \citep{2021Mcca}.

\cite{2018McGuire}  and \cite{2021Mcca} propose one of the two following arguments to explain this. Firstly, it's conceivable that a significant portion of N might not be present primarily in the NH form, whose reaction with butadiene is considered a significant route leading to the formation of pyrrole. Another potential scenario is that the process of cyanation effectively transforms pre-existing aromatic substances, yielding c-C$_{5}$H$_{5}$CN and other rings with CN-functionalization. The formation of 3-pyrroline from pyrrole depends on two successive H-addition reactions at pyrrole's carbon-sites adjacent to the N atom where there is a slight possibility of a H$_{2}$ abstraction process leading to the formation of pyrrole itself \citep{2021MNRAS.505.3157M}. However, \cite{2021MNRAS.505.3157M} suggests that 2 and 3-pyrroline are promising candidates for future astronomical searches, with 3-pyrroline being the most stable isomer \citep{BOGGS1985271}. This further hints at the potential to search for 3-pyrroline in the ISM, which could shed more light on the N-heterocycles' chemistry. Due to our incomplete understanding of pyrrole formation, it remains plausible that 3-pyrroline could be undergoing alternative chemical routes within the ISM besides its association with pyrrole.

In view of the above, the present work explores an in-depth theoretical investigation of 3-pyrroline in interstellar conditions. This involves a reaction mechanism for the formation of 3-pyrroline from vinyl cyanide (CH$_{2}$CHCN), one of the most readily detected and fundamental nitriles in the ISM. We also report accurate rotational and vibrational spectroscopy of 3-pyrroline to facilitate its astronomical searches in diverse interstellar regions. The rotational spectra are studied at three different temperatures (5, 10, and 100 K) to encompass the range of conditions found in N-heterocycle searches, including both cold sources like TMC-1 and warmer environments like Sgr B2. Given the success of hyperfine-resolved structures in identifying molecules within cold environments, such as CN-PAHs in TMC-1 \citep[e.g.][]{2018McGuire,2021Sci...371.1265M}, we present the hyperfine resolved spectrum of 3-pyrroline at 5 K. The highly accurate IR spectrum of 3-pyrroline, which incorporates anharmonic effects, is presented here in light of the JWST ushering in a new era of unparalleled IR sensitivity. This sensitivity opens the door to potentially detecting single molecules. We further investigate how the vibrational spectra of 3-pyrroline change as the concentration of H$_{2}$O ice increases. This is particularly relevant because water ice is a dominant component in dense interstellar clouds. The paper is organized as follows. Section 2 outlines the computational and theoretical methods employed. Section 3 presents the results along with their astrophysical implications, and Section 4 summarizes the conclusions.

\section{Computational Details}
    \label{fig:enter-label}
All the quantum chemical calculations here have been performed using Density Functional Theory (DFT) with Gaussian 16 \citep{g16} suite of programs.


The formation of 3-pyrroline was investigated by employing the B3LYP functional \citep{1988lee,1993JChPh..98.5648B} with Dunning's correlation-consistent basis set augmented with diffuse functions, aug-cc-pVTZ \citep{1989JChPh..90.1007D} and considering dispersion effects-D3(BJ) \citep{grimme2011effect}. We first optimize the reactants and products of each step to calculate the transition states (TS) with the QST2 method \citep{peng1993combining}. The B3LYP/aug-cc-pVTZ, is widely used to study reaction mechanisms in the ISM \citep{2020ApJ...900...85T,2022ApJ...938...15F,10.1093/mnras/stad3770,ijms242316824}. The same methodology has been employed to calculate reaction enthalpies and Gibbs free energies. Additionally, we perform single-point energy calculations for the TS and the optimized reactant structures at the CCSD(T)/aug-cc-pVTZ level to correct the barriers more accurately. Frequency analysis confirmed minima and transition states, with TS geometries connecting minima explored via intrinsic reaction coordinates (IRC).
\begin{figure*}
\centering
    \includegraphics[width=0.6\textwidth]{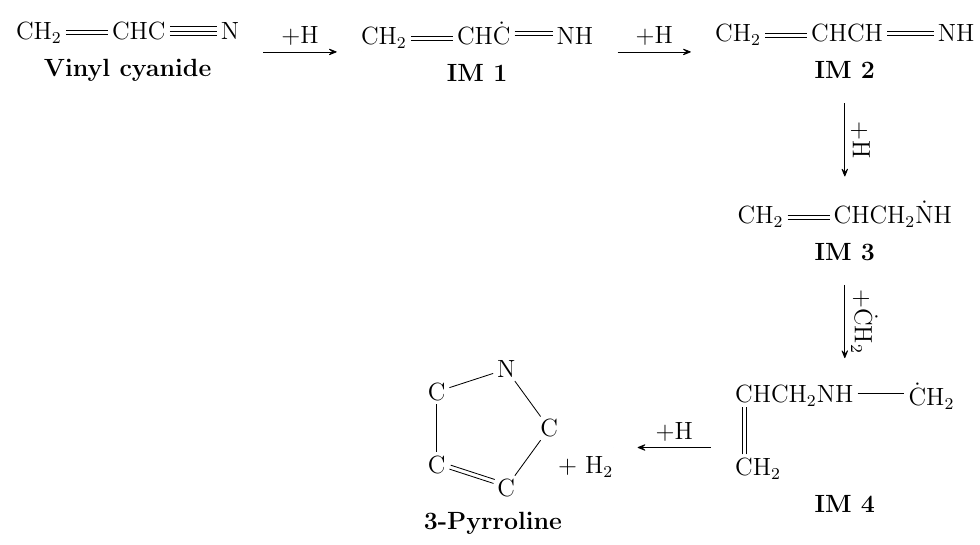}
    \caption{Formation of 3-Pyrroline from vinyl cyanide.}
    \label{fig:2}
\end{figure*}
To mimic the characteristics of ice, molecules in the reaction network are embedded in a continuous solvation field to simulate local effects. This is achieved using the default self-consistent reaction field (SCRF) method, specifically the integral equation formalism (IEF) variant of the polarizable continuum model (PCM), with water acting as the solvent \citep{1997JChPh.107.3032C,doi:10.1021/cr9904009}. Studies have found that an implicit solvation model is preferable to mimic the water ice environments. While explicit water molecules can create a more realistic (inhomogeneous) field, the uncertainty regarding the typical size of interstellar water clusters (i.e., the number of water molecules required) led us to prefer using the SCRF method \citep{2002ApJ...571L.177W,2021ApJ...922..194M}. The implicit solvation model places the molecule inside a cavity within a uniform dielectric medium, like water, representing the solvent. This method works well for simulating ices. The PCM accounts for the reaction field of bulk ice. The reaction fields issuing from a dielectric constant of 78 for water and about 100 for ice \citep{2011JPCA..115.5745A} are very similar. Thus, the solid-phase calculations and thermochemical parameters here are similar to those in ice.

Binding energies (BEs) determine the
residence time of molecules on grain surfaces and are crucial for
astrochemical models to accurately predict molecular abundances in
the ISM \citep{He_2016,WAKELAM201722,Gorai_2017,10.3389/fspas.2021.645243}. Therefore, we have calculated the BEs of reactants, intermediates, and products involved in the reaction using a water cluster of tetramer size as a substrate due to its effectiveness in explaining chemical models \citep{2018das}. Here, BEs are computed at the B3LYP-D3(BJ)/aug-cc-pVTZ level of theory to examine the comparative BEs of reactants, intermediates, and products with water ices, using the following equation.



\vspace{0.4cm}
\hspace{=1.5cm}BE = (\textit{E}$_{substrate}$ + \textit{E}$_{species}$) -- \textit{E}$_{ads}$.
\vspace{0.3cm}

Where, \textit{E}$_{substrate}$ and \textit{E}$_{species}$ are the optimized energies of the grain substrate and species, respectively, where both are obtained separately. \textit{E}$_{ads}$ is the total calculated energy of the species adsorbed on the surface. 

Rotational constants (A, B, C) dictate the intricate spectral fingerprints in rotational spectroscopy. For interstellar environments, rotational constants in the ground vibrational state are crucial. These constants primarily consist of equilibrium rotational values with minor corrections due to vibrational effects \citep[][]{puzzarini2023}.
The equilibrium constants were calculated using the PCS (Pisa Composite Scheme)/Bonds model \citep{doi:10.1021/acs.jpca.3c03999,doi:10.1021/acs.jpclett.3c01380,10.1063/5.0167296} beginning with geometries optimized by the revDSD-PBEP86-D3(BJ) functional combined with the 3F12$^{-}$ basis set \citep{doi:10.1021/acs.jpca.3c06649}. This model incorporates both complete basis set extrapolation and core-valence correlation effects. These inclusions significantly improve the accuracy of calculations for medium- to large-sized molecules, typically by an order of magnitude. The vibrational corrections are acquired by the B3LYP-D3(BJ)/6-31+G* semidiagonal cubic force fields for a good comparison with experimental ground-state rotational constants.
The method delivers highly accurate results for N-heterocycles compared to experiments (better than 0.1\%) like pyridine (within 0.06\% error) and pyrrole (within 0.02\% error) \citep{doi:10.1021/acs.jpca.3c06649}. This accuracy makes PCS/Bonds the ideal choice for studying 3-pyrroline.


Nuclear quadrupole coupling constants and distortion constants were calculated using frozen core approximation coupled cluster method with the perturbative treatment of triple excitations ((fc-CCSD(T)) with the aug-cc-pVTZ basis set in CFOUR \citep{2020JChPh.152u4108M}. This method has been extensively used for the calculation of spectroscopic constants \citep{2023PCCP...25.1421P,doi:10.1021/acs.jpca.3c05741}.
The calculation of nuclear quadrupole coupling constants, $\chi_{ij}$, requires the calculation of the electric field gradient at the
quadrupolar nucleus

\vspace{0.3cm}
\hspace{0.6cm}$\chi_{ij}$ = eQ $\times$ $q_{ij}$
\vspace{0.3cm}

here, \textit{i} and \textit{j} denote the inertial axes, eQ stands for the nuclear quadrupole moment, and $q_{ij}$ denotes \textit{ij}-th element in the electric field gradient tensor \citep{gordy1953microwave,2010IRPC...29..273P}.

We have utilized the PGOPHER general purpose software \citep{western2017} to simulate the rotational spectra. All spectra have been simulated with rotational-vibrational coupling, anharmonic, quartic centrifugal-distortion, and nuclear quadrupole coupling corrections at the Vibrational second-order perturbation theory (VPT2) level using Watson's S-reduced asymmetric rotor Hamiltonian \Romannum{1}$^{\rm r}$ axis representation. 
Pickett's SPCAT \citep{pickett1991} program was used to predict transitions for the target molecules, accompanied by estimated errors, and presented them in the JPL standard catalog format. This format facilitates convenient searches for the
considered molecule in space.

VPT2 utilizes quadratic, cubic, and quartic force constants to compute anharmonic vibrational frequencies for polyatomic molecules. Additionally, a new generalized version (GVPT2) has been developed to variationally handle near-resonant terms, which is particularly needed for the molecules having CH and NH stretching regions in their IR spectra \citep{2005JChPh.122a4108B,2010CPL...496..157B,2012JChPh.136l4108B}. Within the GVPT2 model, vibrational anharmonicities, mode couplings, and resonances can be efficiently addressed for both energy levels and intensities \citep[e.g.][]{2021FrASS...8...77Y}. In this work, the anharmonic IR absorption spectra of 3-pyrroline were computed utilizing GVPT2 using B3LYP-D3(BJ) with N07D basis set \citep{2008CPL...454..139B} with very tight optimization and a superfine grid. The B3LYP-D3(BJ) method is typically used for examining weakly bonded systems to understand the influence of weak interactions on anharmonic IR spectra. The B3LYP-D3(BJ)/N07D method yields highly accurate results for anharmonic IR spectra of aromatic molecules, with the agreement between theory and experiment within 1\% reducing to 0.1\% for CH stretching region in PAHs \citep{2018PCCP...20.1189M}. The GVPT2 approach in Gaussian 16 has been proven accurate for addressing anharmonicity in the IR spectra of PAH-related molecules, as demonstrated with pure and nitrogen-containing PAHs (carbon atoms $\ge$ 22) \citep{2018ApJS..238...18C,2021ApJ...923..238L}. In general, GVPT2 includes a dual-phase process: initially, resonant terms are singled out using an ad hoc test \citep{1995JChPh.103.2589M} and subsequently eliminated, which is termed as deperturbed VPT2 (DVPT2). In the second phase, the eliminated terms are reintroduced via a variational approach. This method is acknowledged for yielding highly precise outcomes \citep{1995JChPh.103.2589M,2014PCCP...16.1759B,2015JChPh.143v4314M,2016JChPh.145h4313M,2018ApJS..238...18C,2021ApJ...923..238L,2021FrASS...8...77Y,2021JPCA..125.1301F, 2024JChPh.160k4312E}. For example, the majority of the anharmonic wavenumbers computed with GVPT2 for pyrrole (C$_{4}$H$_{5}$N), a molecule similar to 3-pyrroline (C$_{4}$H$_{7}$N), align with experiment within 5 cm$^{-1}$ \citep{2014PCCP...16.1759B}.

As H$_{2}$O molecules constitute a substantial fraction of interstellar ice mantles within dense interstellar environments, we have also focused on examining the infrared features of 3-pyrroline in the presence of increasing concentrations of H$_{2}$O. Specifically, we investigated how varying concentrations of water impact the fundamental absorption bands of 3-pyrroline. To explore these effects, we computed the IR spectra of 3-pyrroline in the presence of increasing numbers of H$_{2}$O monomers, modeling the interaction with water clusters. The structures of the 3-pyrroline water ices investigated here were thoroughly optimized, achieving the three-dimensional arrangement of atoms that minimizes the total energy. Harmonic frequencies were then computed using DFT in conjunction with the B3LYP/aug-cc-pVTZ level of theory. Compared to lower-level basis sets, the aug-cc-pVTZ basis set offers improved accuracy in describing electron-correlation effects, leading to better agreement with experimental data \citep[e.g.][]{2021ApJ...917...68L}. The concentrations of H$_{2}$O and 3-pyrroline (C$_{4}$H$_{7}$N:H$_{2}$O) are selected as 1:1, 1:2, 1:3, 1:4, 1:8, 1:10, and 1:16, respectively. These concentrations are chosen to investigate the systematic effect of increasing H$_{2}$O concentrations on fundamental IR bands of 3-pyrroline.

\begin{figure*}
\begin{subfigure}{0.5\textwidth}
\includegraphics[width=1.0\textwidth]{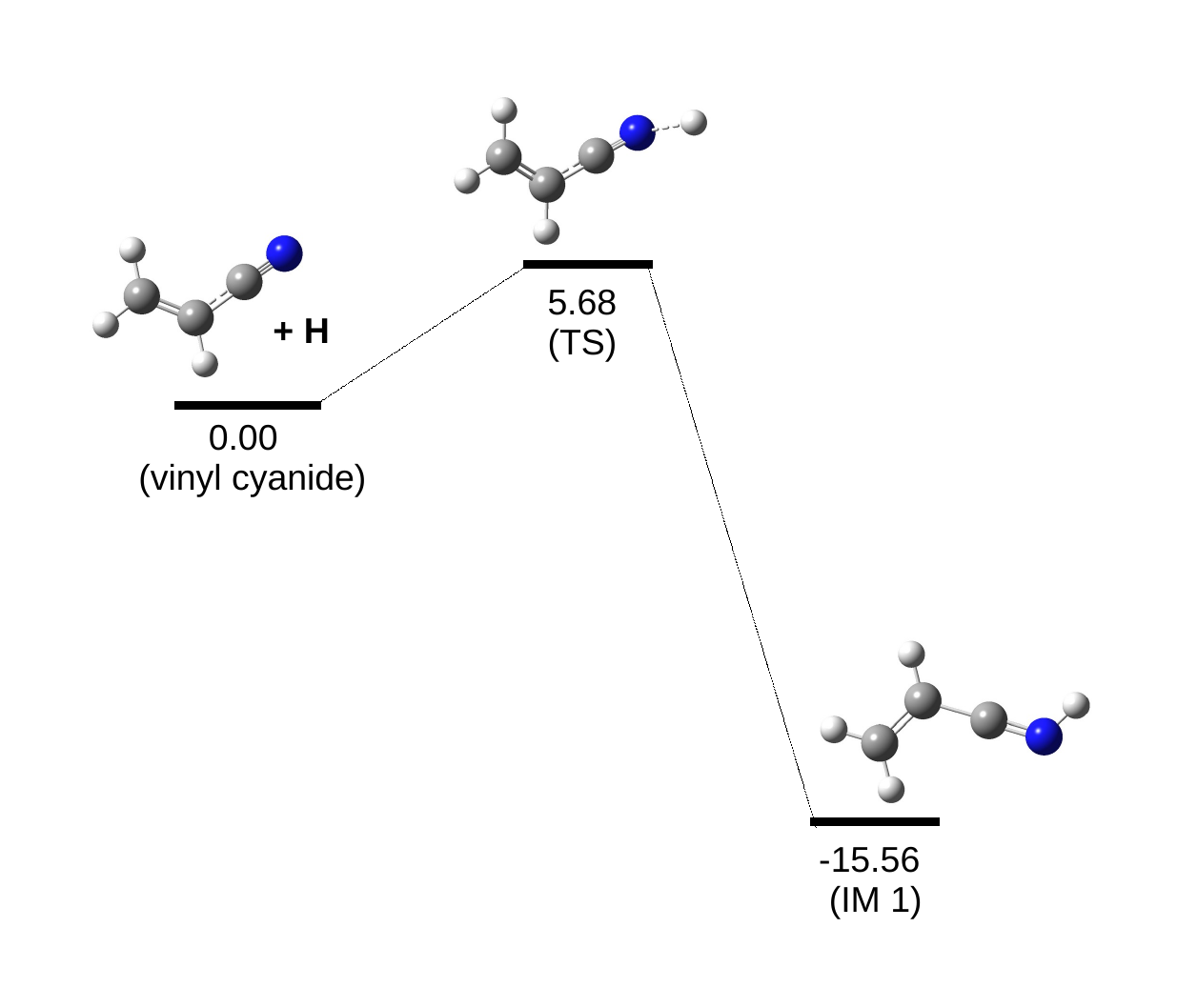}
\caption{(a)}
\label{fig1}
\end{subfigure}%
\begin{subfigure}{0.5\textwidth}
\includegraphics[width=1.0\textwidth]{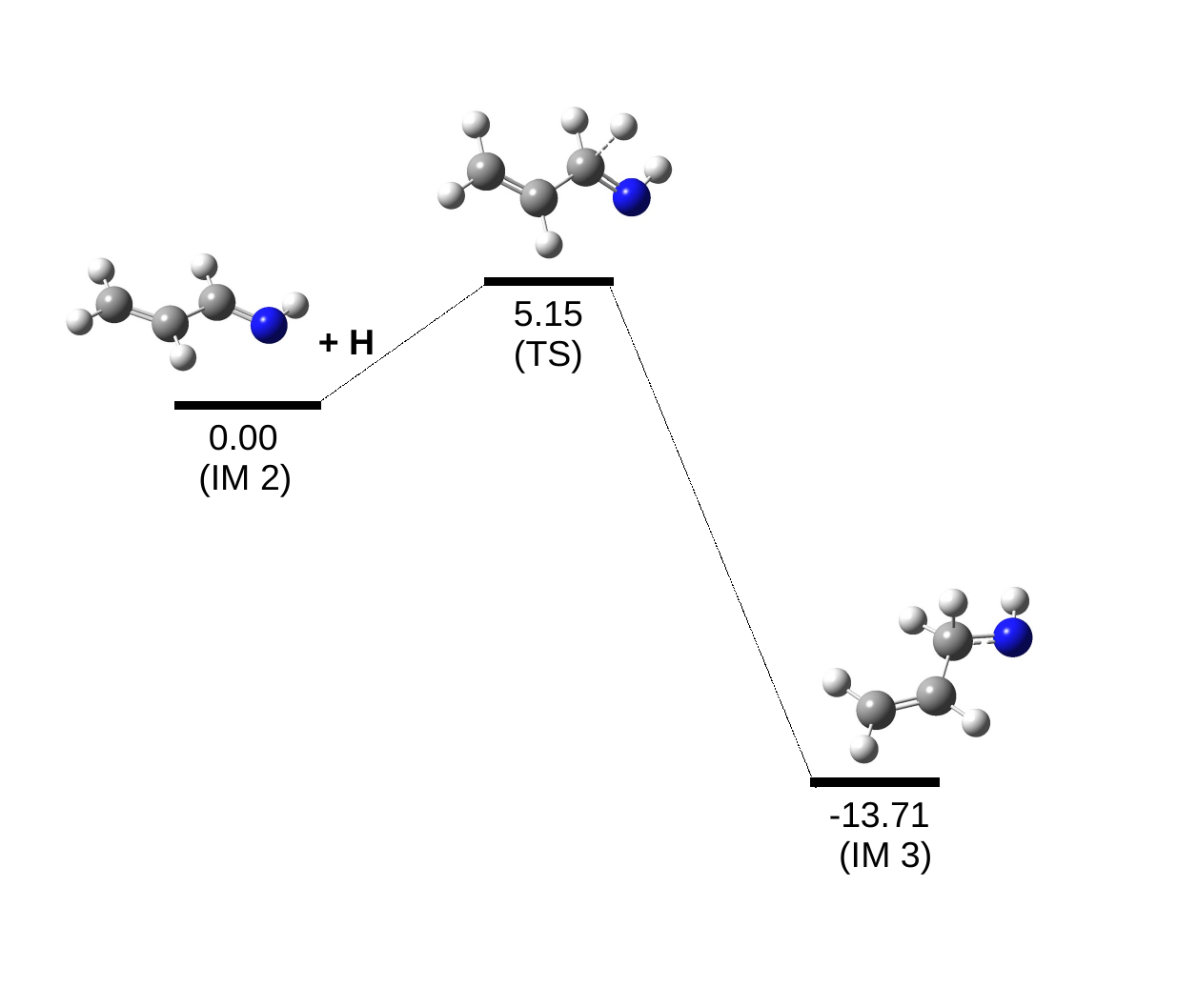}
\caption{(b)}
  \label{fig2}
\end{subfigure}
\caption{Potential energy diagram for (a) reaction 1, and (b) reaction 3; calculated using CCSD(T)/aug-ccpVTZ level of theory reported in kcal/mol.}
\end{figure*}

\begin{table*}
\begin{threeparttable}
	\centering
	\caption{The ZPE corrected activation barrier at the CCSD(T)/aug-ccpVTZ level of theory, reaction enthalpy and gibbs free energies at the B3LYP/aug-ccpVTZ level of theory.}
	\label{tab:example_table}
\begin{tabular}{|c|c|c|c|c|}
		\hline
	  Reaction number & Reaction Path & Calculated  & Enthalpy & Gibbs Free Energy \\
        (type) &  &  Activation barrier (kcal/mol) & (kcal/mol) &(kcal/mol) \\
		\hline
		1 (NR) &   CH$_{2}$CHCN + H  $\longrightarrow$ CH$_{2}$CH\.{C}NH &5.68&-21.39 & -14.38 \\
		2 (RR) &CH$_{2}$CH\.{C}NH + H $\longrightarrow$ CH$_{2}$CHCHNH&-&-95.59&-86.75\\ 
		3 (NR) &CH$_{2}$CHCHNH + H $\longrightarrow$ CH$_{2}$CHCH$_{2}$\.{N}H&5.15&-25.28&-18.47\\
		4 (RR) &CH$_{2}$CHCH$_{2}$\.{N}H + \.{C}H$_{2}$$\longrightarrow$ CH$_{2}$CHCH$_{2}$NH\.{C}H$_{2}$ &-&-105.56 &-94.17\\
	    5 (NR) 
    &CH$_{2}$CHCH$_{2}$NH\.{C}H$_{2}$ + H $\longrightarrow$ \.{C}HCHCH$_{2}$NH\.{C}H$_{2}$ + H$_{2}$  &---&---&---\\
		\hline
		 
\end{tabular}
\begin{tablenotes}
	\item Notes: NR and RR refer to neutral-radical and barrierless radical-radical reactions, respectively. 
\end{tablenotes}
\end{threeparttable}
\end{table*}

\begin{table*}
\begin{threeparttable}
	\centering
	\caption{Calculated binding energies of reactant, intermediates, and product with water tetramer at the B3LYP/aug-ccpVTZ level of theory.}
	\label{tab2}
\begin{tabular}{|c|c|c|c|}
		\hline
	  Serial number & Species & Ground state  & Binding energy \\
         & & &  H$_{2}$O Tetramer (K)\\
		\hline
		1 (Vinyl cyanide) &   CH$_{2}$CHCN  &Singlet&1294 \\
		2 (IM 1) &CH$_{2}$CH\.{C}NH&Doublet&3337\\ 
		3 (IM 2) &CH$_{2}$CHCHNH&Singlet&834\\
		4 (IM 3) &CH$_{2}$CHCH$_{2}$\.{N}H &Doublet&722 \\
            5 (IM 4) &CH$_{2}$CHCH$_{2}$NH\.{C}H$_{2}$  &Doublet&573 \\
            6 (3-Pyrroline) &CHCHCH$_{2}$NHCH$_{2}$ &Singlet&6884\\ 
        \hline
		 
\end{tabular}
\end{threeparttable}
\end{table*}
\section{Result and discussion}
\subsection{Probable formation of 3-pyrroline}
In the ISM, smaller aromatic molecules with one or two rings, including N-heterocycles, might form locally from simpler building blocks (bottom-up) \citep[e.g.][]{2021SciA....7.4044D,2021MNRAS.505.3157M,2021McCar,2022JPCA..126.2716B}. This is supported by the recent discoveries of CN-PAHs and a pure PAH (indene) in dense molecular clouds like TMC-1 \citep{2018McGuire,2021Sci...371.1265M,2021ApJ...913L..18B,2023A&A...677L..13A}. Additionally, smaller aromatic molecules (less than 20-30 carbon atoms) likely cannot survive the harsh journey from the diffuse ISM to cold molecular clouds due to destruction by ultraviolet radiation \citep{2020ApJ...888...17C}. Therefore, the current section explores the possibility of the formation of 3-pyrroline from vinyl cyanide, which is one of the most abundant and readily identified cyanide in cold, dense molecular clouds as it was detected in various diverse environments of the ISM \citep[e.g.][]{1975ApJ...195L.127G,1997ApJS..108..301S,1981Natur.293...45C,1983ApJ...267L..53M,2008Ap&SS.313..229A}. The high abundance of hydrogen atoms in dense molecular clouds makes H-addition and H$_{2}$-abstraction reactions especially significant, likely playing a key role in the formation of 3-pyrroline from vinyl cyanide. Following this, a series of five steps of chemical reactions for the formation of 3-pyrroline is shown graphically in Figure 2, and the energetics of the reactions are listed in Table 1 with activation barriers, reaction enthalpies, and Gibbs free energies (all ZPE corrected). The intermediate products formed in each step here are depicted as IM 1--IM 4 in Figure 2. The potential energy diagram
of these reactions is shown in Figure 3 (a) and Figure 3 (b). The enthalpies ($\Delta$H) and Gibbs free energies ($\Delta$G) possess negative values (Table 1), which shows the exothermic and spontaneous nature of all the reactions. 

While H-addition to the carbon atom of the double bond is more viable in vinyl cyanide, the recent detection of protonated vinyl cyanide \citep{10.1063/1.4793316} suggests that H-addition at the nitrogen site might also occur under specific conditions, particularly on dust grain surfaces in cold ISM regions. This forms the radical CH$_{2}$CH\.{C}NH (IM 1) with an achievable barrier of 5.68 kcal/mol. This newly formed radical can then react with an H atom without any further barrier, leading to a closed-shell molecule CH$_{2}$CHCHNH (allylimine; IM 2). Allylimine has been tentatively detected in the G+0.693-0.027 molecular cloud \citep{2023A&A...669A..93A}, and in interstellar conditions, imines are also considered as reactive species on dust grain surface \citep[e.g.][]{puzzarini2023}. 
Further hydrogenation of allylamine at C (adjacent to N) would lead to a radical intermediate CH$_{2}$CHCH$_{2}$\.{N}H (IM 3) with a barrier of 5.15 kcal/mol. The fourth intermediate product, CH$_{2}$CHCH$_{2}$NH\.{C}H$_{2}$ (IM 4), is formed through radical-radical reaction when \.{C}H$_{2}$ present in the dense molecular clouds \citep{1995ApJ...438..259H,2005A&A...431..203P} reacts at the radical N site. The final step might involve the abstraction of an H atom followed by ring closure, ultimately leading to the formation of 3-pyrroline. However, this final stage remains enigmatic, necessitating additional investigations. It is worth noting that the double hydrogenation of pyrrole to form 3-pyrroline also presents uncertainties. Nevertheless, the existing energetic considerations here suggest a plausible route from vinyl cyanide. The challenge lies in the absence of identified H abstraction partners, prompting the exploration of alternative interstellar reactants like OH to unravel this intriguing pathway further.

The BEs of the reactants, intermediates, and products have been computed and presented in Table 2. This allows for a comparison of the residing time of the molecules and their credibility in the formation reaction on dust grain surfaces as the diffusion of a particle on the grain surfaces is a fraction of its BE \citep{2017SSRv..212....1C}.
Since water (H$_{2}$O) makes up the predominant portion ($\sim$70$\%$ by mass) of a grain mantle \citep[e.g.][]{10.1111/j.1365-2966.2010.17343.x,2021A&A...648A..24V,2021FrASS...8...78D}, we have calculated the BEs using water tetramers as a substrate due to their efficiency in explaining certain chemical models \citep{2018das}. The calculated BEs should exhibit a variety of values depending upon the various binding sites, so the average of BE values have been taken and provided in Table 2 \citep{2017EPJD...71..340S,Das_2018,Ferrero_2020}. Ground state spin multiplicity of the molecules is also noted (Table 2). As shown in Table 2, there is a decrease in the values (from IM 2 to IM 4: 834 K, 722 K, and 573 K, respectively), suggesting easier diffusion of intermediates on the surface, thus promoting the reaction's feasibility. Notably, 3-Pyrroline has the highest binding energy (6884 K), highlighting its potential residence on the grain surface.

\begin{table}
\caption{Spectroscopic parameters (MHz) in the ground state of 3-pyrroline.}
    \label{tab:be}
    \centering
    \begin{threeparttable}
    \begin{tabular}{lcc}
    \hline
\hline
     &  3-Pyrolline\\\hline
    
Parameters&B3LYP\\
&aug-cc-pVTZ\\
   \hline
A$_{0}$$^{\rm a}$ & 7698.6674\\
B$_{0}$ & 7630.0130\\
C$_{0}$ & 4062.8761\\ 
D$_{\rm J}^{\rm b}$ $\times$10$^{6}$ & 893.0330\\
D$_{\rm JK}$ $\times$10$^{3}$ & 1.3258\\
D$_{\rm K}$ $\times$10$^{3}$& 0.2678\\
d$_{\rm 1}$ $\times$10$^{6}$ & -427.7390\\
d$_{\rm 2}$ $\times$10$^{6}$ & -140.9430\\
$\chi_{aa}^{\rm c}$  & 3.664\\
$\chi_{bb}$  & 1.630\\
$\chi_{cc}$  & -5.295\\
($\chi_{bb}$--$\chi_{cc}$) & 6.925\\
$\mu^{\rm d}$ (Debye) & $\mu_{\rm a}$ = -1.1844\\
& $\mu_{\rm b}$ = -0.3662\\
& $\mu_{\rm c}$ = 0.0000\\
& $\mu_{\rm tot.}$ = 1.5506\\
\hline
\end{tabular}
        \begin{tablenotes}
	\item Notes. $^{\rm a}$Ground state Rotational constants at  revDSD-PBEP86-D3(BJ)/3F12$^{-}$ combined with B3LYP-D3(BJ)/6-31+G$^{*}$ for vibrational corrections, $^{\rm b}$Quartic centrifugal distortion constants, $^{\rm c}$$^{14}$N quadrupole coupling constants about the principal inertial axis system, $^{\rm d}$Dipole moment calculated at fc-CCSD(T)/aug-cc-pVTZ. 
\end{tablenotes}
 \end{threeparttable}
\end{table}
\subsection{Rotational spectroscopy}

Table 3 presents the spectroscopic constants of 3-pyrroline including ground state rotational constants (A$_{0}$, B$_{0}$, C$_{0}$), centrifugal distortion constants (D$_{\rm J}$, D$_{\rm JK}$, D$_{\rm K}$, d$_{\rm 1}$, d$_{\rm 2}$), quadrupole coupling constants ($\chi_{aa}$, $\chi_{bb}$, $\chi_{cc}$), and dipole moment ($\mu$). 
\begin{figure}
\includegraphics[width=0.5\textwidth]{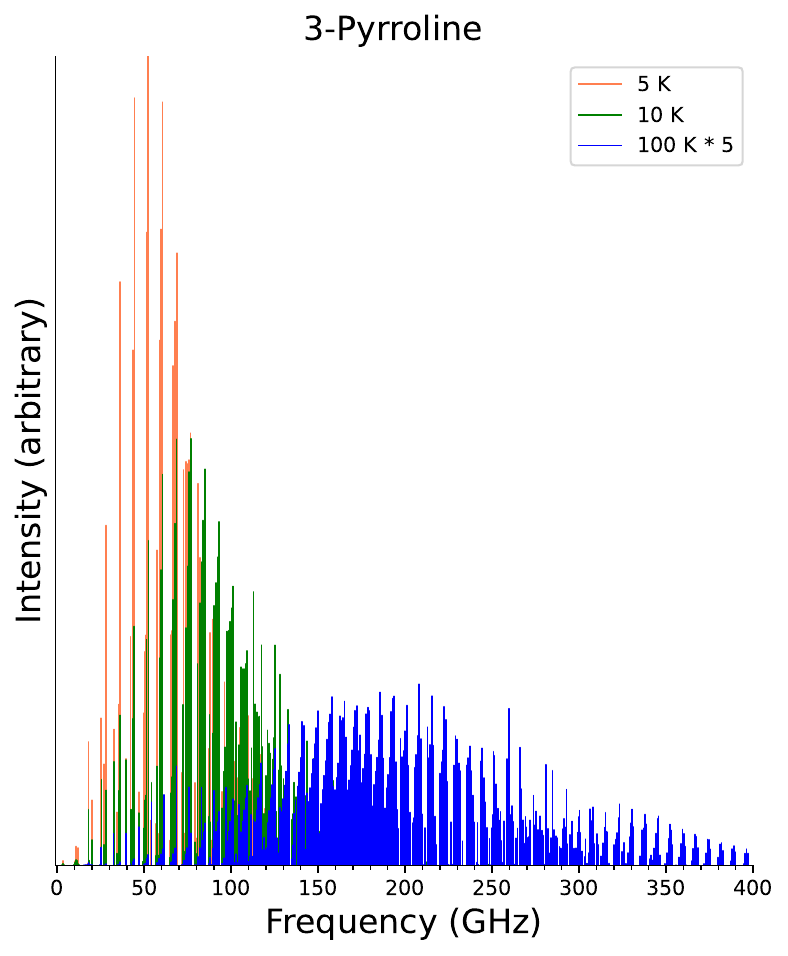}
\caption{Theoretical rotational spectra of 3-pyrroline at 5 K (red), 10 K (green), and 100 K (blue). For display, the relative intensities for the 100 K plot have been multiplied by a factor of 5.} 
\end{figure}
\begin{figure}
\subfloat{\includegraphics[width=0.5\textwidth]{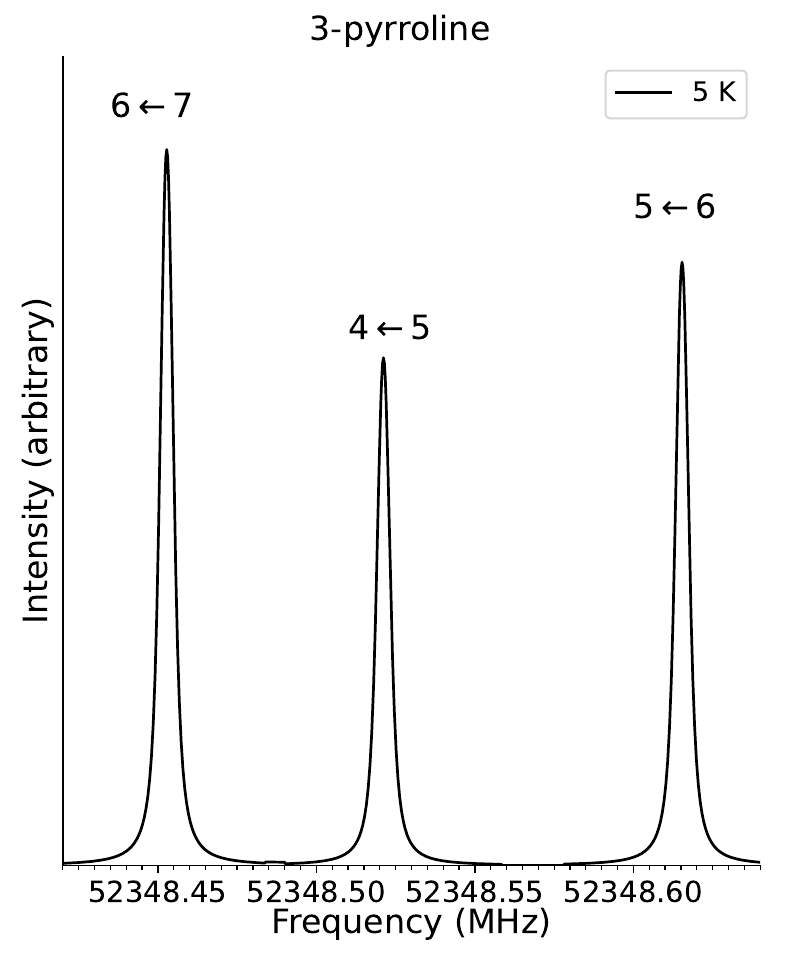}}
\caption{The strongest transition of 3-pyrroline at 5 K (J$'_{K'_a,K'_c}$ $\rightarrow$ J"$_{K"_a,K"_c}$ = 6$_{1,6}$ $\rightarrow$ 5$_{1,5}$) 
showing the resolved hyperfine structure (F" $\rightarrow$ F') in the spectrum.
}
\end{figure}
3-Pyrroline is a near oblate type asymmetric top molecule with $\kappa$ = 0.96, similar to pyrrole ($\kappa$ = 0.94). 
Compared to molecules of similar size, the rotational constants of 3-pyrroline are relatively higher. For instance, 1- and 2-cyano-cyclopentadiene, a five-membered ring molecule detected in the TMC-1 \citep{2021NatAs...5..176M}, have a similar number of atoms as 3-pyrroline (12 atoms) show rotational constants as A$_{0}$$\sim$8235--8350 MHz, B$_{0}$$\sim$1900 MHz, C$_{0}$$\sim$1550 MHz, which are lower than those of 3-pyrroline (Table 3). This makes 3-pyrroline suitable for searching in the ISM in terms of their partition function values expressed as Q$_{\text{rot}}$ =[(kT/h)$^{3}$ ($\pi$/ABC)]$^{\frac{1}{2}}$, which relies on temperature (T) and rotational constants (A, B, and C). The partition function along with dipole moments, is a crucial quantity for a molecule to be detected by radio telescopes. The rotational partition function values (Q$_{\rm rot}$) of 3-pyrroline for seven CDMS (The Cologne Database for Molecular Spectroscopy) standard temperatures as implemented in the JPL database \citep{1998JQSRT..60..883P} along with an additional temperature at 5 K are given in Table 4.
The centrifugal distortion and quadrupole coupling constants for 3-pyrroline, given in Table 3, are computed at the fc-CCSD(T)/aug-cc-pVTZ level of theory, which has been reported to perform significantly better compared to experiments (within $\sim$4-5$\%$) for organic molecules \citep[e.g.][]{doi:10.1080/00268976.2021.1955988,https://doi.org/10.1002/jcc.27283}.
\begin{table}
\centering
\begin{threeparttable}

    \caption{Rotational partition function values (Q$_{\text{rot}}$) of 3-pyrroline at given temperatures.}
    \begin{tabular}{lccc}
    \hline
    \hline
    Temperatures (K) & Q$^{*}$$_{\text{rot}}$\\
    \hline
    300.000   &  28368.4076\\    
    225.000   &  26058.7461 \\ 
    150.000   &  22113.3889 \\   
     75.000   &  14120.5241  \\ 
     37.500   &  6866.4491  \\ 
     18.750   &   2650.6082  \\ 
      9.375    &   946.3981   \\
      5.000   &    370.8478   \\
\hline
\end{tabular}
\begin{tablenotes}
\item $^{*}$Q$_{\text{rot}}$=[(kT/h)$^{3}$($\pi$/ABC)]$^{\frac{1}{2}}$, mainly depends on temperature (T) and rotational constants (A,B,C). 
\end{tablenotes}
\end{threeparttable}
\end{table}
For dipole moments, 3-pyrroline shows a viable polarity with $\mu_{\rm tot}$=1.5 D but lower compared to CN-PAHs, which is a common behavior among N-heterocycles with an N atom inside the structure and molecules containing a CN side group \citep{KISIEL2003115,10.1093/mnras/sty557,2019JMoSp.36311175V,10.1093/mnras/stac3157,2023PCCP...2519066V}. 
By symmetry, 3-pyrroline has both a-and b-type rotational transitions, with the b-type transition much weaker than the a-type (with $\mu_{\rm a}$ = --1.1844
\& $\mu_{\rm b}$ = --0.3662).


Using the accurate rotational, centrifugal distortion, and nuclear quadrupole coupling constants from Table 3, the simulated rotational spectrum of 3-pyrroline is shown in Figure 4. The rotational spectra in Figure 4 are provided at 5 K (red), 10 K (green), and 100 K (blue) temperatures to account for colder and warmer temperature regions in the ISM. It should be noted that N-heterocycles have been primarily searched in colder regions \citep{2018McGuire,2021McCar,2022JPCA..126.2716B}. The intensity at 100 K is multiplied by a factor of 5, respectively, to account for the influence of a larger partition function that reduces the strength of the transitions with increasing temperature. The most intense rotational transitions of 3-pyrroline at some warmer temperatures fall in the millimeter (110–300 GHz) band, specifically near 200 GHz (215363.8940 MHz). On the other hand, at colder temperatures (at 10 K and 5 K), the transitions having maximum peak intensities lie within the V (40--75 GHz) band, at around 68 GHz (68590.1327 MHz) for 10 K and at around 60 GHz (60470.0734 MHz) for 5 K. 
\begin{table}
\centering
    \caption{Hyperfine resolved strongest transitions of 3-pyrroline at 5 K.}
    \begin{tabular}{ccc}
    \hline
    \hline
    \multicolumn{3}{c}{5 K}\\
    \hline
      $J^{'}_{K^{'}_{a},K^{'}_{c}}$ $\to  J_{K_{a},K_{c}}$ & F"$\rightarrow$F'  &Calculated\\
     &&Frequency (MHz)\\
      \hline
$6_{1,6}$$\to5_{1,5}$& 7$\rightarrow$6 & 52348.4528\\
& 5$\rightarrow$4&52348.5212\\
& 6$\rightarrow$5&52348.6155\\
       
        \hline
       \end{tabular}
    \label{tab:3}
\end{table}
Due to the importance of hyperfine resolved spectra in the detection of molecules in colder ISM regions, such as recently detected CN-PAHs in the TMC-1\citep[for e.g.][]{2018McGuire}, the hyperfine spectra for the strongest transition of 3-pyrroline have been calculated at 10 K, and shown in Figure 5 and transition frequencies are given in Table 5. For 3-pyrroline: the strongest transition at 10 K lies at J'=6 around 52.3 GHz in the V band (40--75 GHz) (Figure 5 and Table 5). 

Based on the above, the current rotational spectroscopic analysis suggests that 3-pyrroline is a promising candidate for detection in the ISM, specifically in colder regions. The accurate rotational data and lines provided here are intended to guide future laboratory and observational studies. This work concludes that 3-pyrroline could be searched around 52 GHz in colder ISM regions. Further high-resolution experiments are needed to support its radio astronomical detection.    
\begin{figure*}
\centering
  \includegraphics[width=0.45\linewidth]{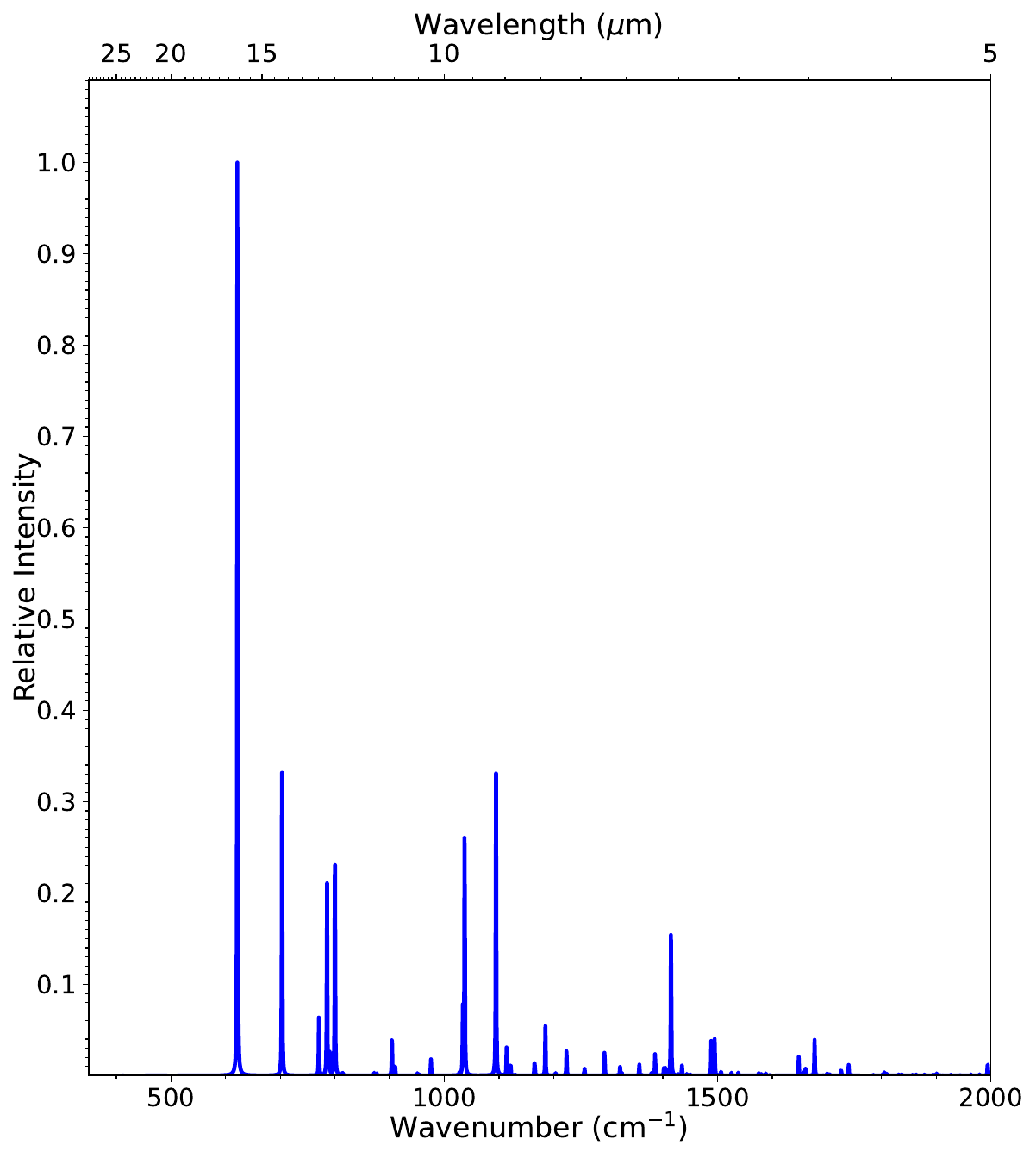}
  \includegraphics[width=0.45\linewidth]{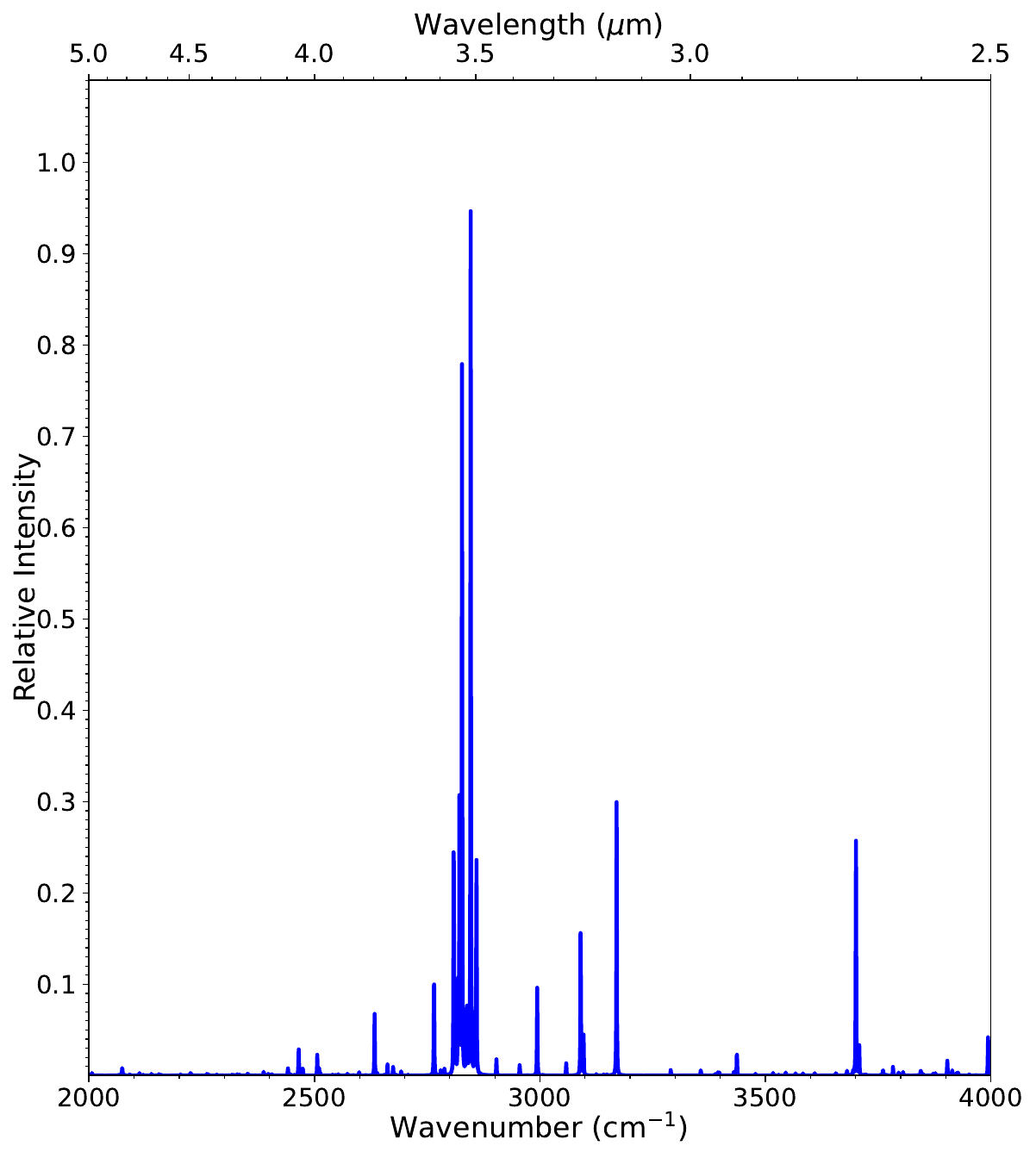}
\caption{Anharmonic gas phase IR absorption spectra of 3-pyrroline with an FWHM of 1 cm$^{-1}$ and a Lorentzian line profile.}
\label{fig:fig}
\end{figure*}
\subsection{IR spectra of 3-pyrroline} 
\begin{table}
    \centering
    \begin{threeparttable}
      \caption{Anharmonic IR frequency (cm$^{-1}$ and $\mu$m), intrinsic intensity (km mol$^{-1}$), relative intensity (I$_{\rm rel.}$), and mode number of 3-pyrroline.}
    \begin{tabular}{ccccc}\hline\hline
     \multicolumn{2}{c}{Frequency}   & Intensity & I$_{\rm rel.}$ & Mode \\
        (cm$^{-1}$) & ($\mu$m) & (km mol$^{-1}$) & & \\ \hline
        621.38 & 16.09 &79.42 & 1.00 & $\nu_{28}$\\
        703.09 & 14.22 &28.42 & 0.36 & $\nu_{27}$\\
        770.55 & 12.97 &05.10 & 0.06 & $\nu_{30}$ + $\nu_{28}$\\
        785.43 & 12.73 &17.56 & 0.22 & $\nu_{30}$ + $\nu_{27}$\\ 
        799.86 & 12.50&22.26 & 0.28 & $\nu_{25}$\\
        904.25 & 11.05&03.81 & 0.04 & $\nu_{24}$\\
        1037.17 & 9.64&23.16 & 0.29 & $\nu_{19}$\\
        1094.56 & 9.13&27.27 & 0.34 & $\nu_{18}$\\
        1184.92 & 8.43&04.50 & 0.05 & $\nu_{15}$\\
        1414.85 & 7.06&13.82 & 0.17 & $\nu_{11}$\\
        1494.50 & 6.68&03.10 & 0.04 & $\nu_{27}$ + $\nu_{22}$\\
        2633.27 & 3.79&05.79 & 0.07 & $\nu_{14}$ + $\nu_{12}$\\ 
        2765.11 & 3.61&09.92 & 0.12 & $\nu_{12}$ + $\nu_{11}$\\ 
        2808.72 & 3.56&23.81 & 0.29 & $\nu_{6}$\\ 
        2821.56 & 3.54&24.39 & 0.30 & 2$\nu_{11}$\\ 
        2827.16 & 3.53&61.59 & 0.77 & $\nu_{7}$\\ 
        2846.35 & 3.51&75.84 & 0.95 & $\nu_{4}$\\ 
        2859.15 & 3.49&22.46 & 0.28 & $\nu_{5}$\\ 
        2993.93 & 3.34&07.63 & 0.09 & $\nu_{10}$ + $\nu_{9}$\\ 
        3090.27 & 3.23&14.93 & 0.18 & $\nu_{2}$\\ 
        3096.58 & 3.22&04.13 & 0.05 & $\nu_{3}$\\ 
        3170.06 & 3.15&23.27 & 0.29 & $\nu_{30}$ + $\nu_{1}$\\ 
        3700.83 & 2.70&22.13 & 0.27 & $\nu_{28}$ + $\nu_{1}$\\ \hline
     \end{tabular}
    \label{tab:my_label}
\begin{tablenotes}
\item Only modes exceeding 0.04 relative intensity are reported. 
\end{tablenotes}
    \end{threeparttable}
\end{table}
\begin{figure}
\centering
  \includegraphics[width=\columnwidth]{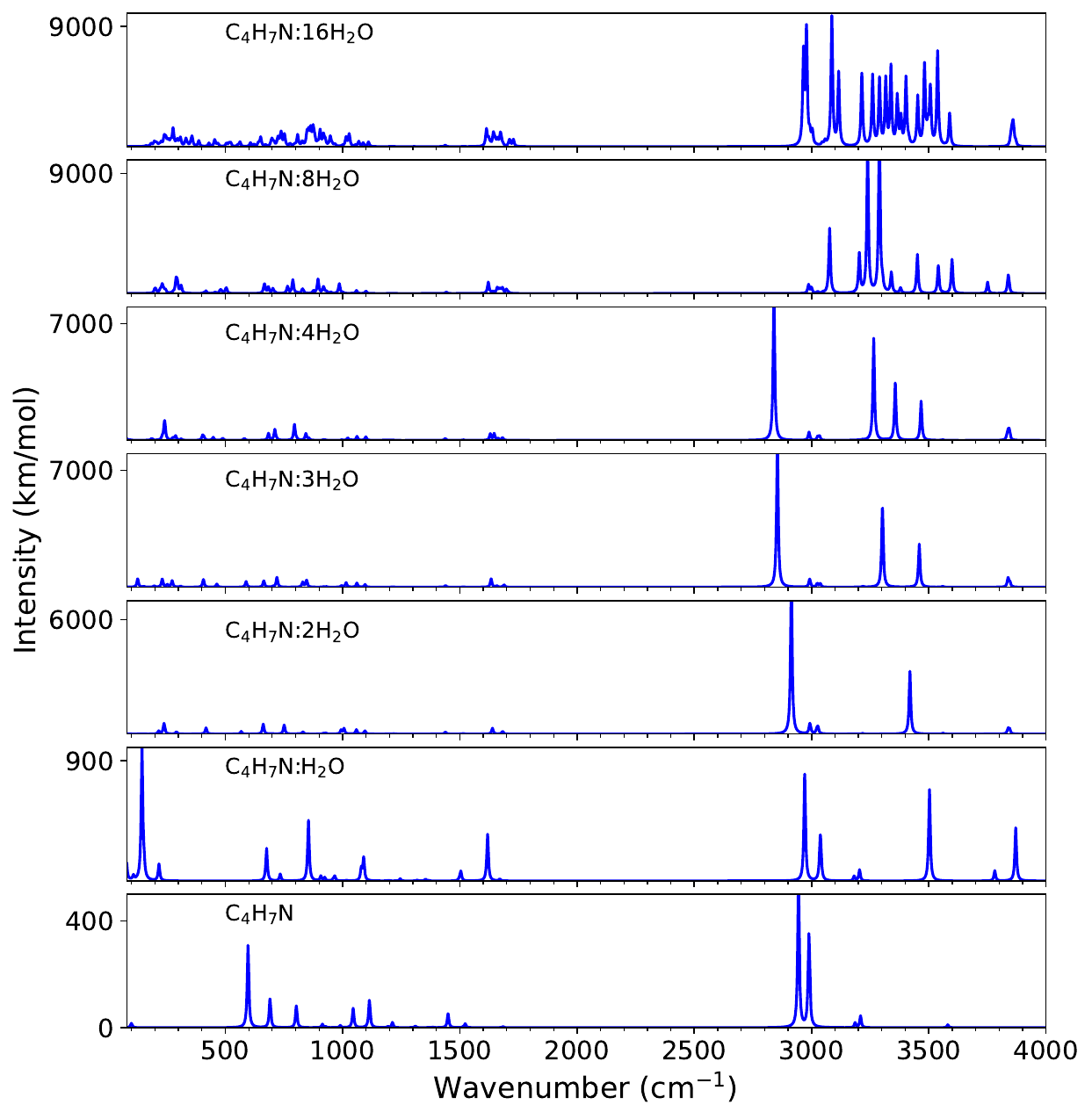}
\caption{Gas phase harmonic IR absorption spectra of 3-pyrroline with different H$_{2}$O concentrations using an FWHM of 8 cm$^{-1}$ and a Gaussian line profile.}
\label{fig:fig}
\end{figure}

There are several IR lines that indicate the presence of PAHs and PAH derivatives in dense molecular clouds \citep{2011A&A...525A..93B}. N-heterocycles, such as 3-pyrroline, are likely to freeze significantly on the ice mantles of dust grains in cold and dense ISM, and their presence can be detected through IR absorption features. For example, a weak band around 3.25 $\mu$m due to CH stretch vibrations \citep[e.g.][]{1995ApJ...449L..69S}, a widely observed feature at 3.47 $\mu$m attributed to CH stretching in CH$_{2}$ groups \citep[e.g.][]{2021ApJ...908..239C}, CC stretching and CH out-of-plane bending vibrations at 6.2 $\mu$m \citep{2001A&A...376..254K}, and 11.2 $\mu$m \citep{2000ApJ...544L..75B}, respectively, may also contribute to IR absorption features in dense ISM. Since aromatics have been detected in the gas-phase, the accurate IR absorption spectra of 3-pyrroline in gas-phase are depicted in Figure 6, convolved with an FWHM of 1 cm$^{-1}$ and Lorentzian profile for detailed analysis of spectral lines. Due to strong resonances effects in the C-H and N-H stretching regions ($\sim$2500--4000 cm$^{-1}$), the spectrum is plotted in two ranges: $\sim$500--2000, and 2000--4000 cm$^{-1}$ with relative intensities (Figure 6).
Considering that a significant portion of interstellar dust exists in the form of water (H$_{2}$O) ices, IR spectra of 3-pyrroline in the presence of water ices are studied and illustrated in Figure 7, convolved with an FWHM of 8 cm$^{-1}$ and Gaussian profile for a continuous spectra. The anharmonic IR frequencies of gas-phase 3-pyrroline are listed in Table 6, along with intensities at both absolute and relative scales. 

In Table 6, the most intense IR features of 3-pyrroline are the $\nu_{28}$ and $\nu_{4}$ fundamentals present at 621.38 cm$^{-1}$ (16.09 $\mu$m) and 2846.35 cm$^{-1}$ (3.51 $\mu$m), respectively. The dominant vibrations are NH out-of-plane bending at 16.09 $\mu$m and CH stretching in CH$_{2}$ groups at $\sim$3.50 $\mu$m. Furthermore, the $\nu_{28}$ fundamental at 16.09 $\mu$m lies in a part of the long-wavelength IR spectrum that is relatively free of features from large PAHs \citep{2004ASPC..309..141P} and silicates \citep{2001A&A...372..165M}, making 3-pyrroline a feasible target to search for with JWST or other IR facilities. 
The observed 3.47 $\mu$m absorption feature is primarily attributed to ammonia hydrates \citep{2001A&A...365..144D,2002A&A...394.1057D}, though other sources have been proposed, such as CH stretching in CH$_{2}$ groups in PAHs \citep{2021ApJ...908..239C}. 3-Pyrroline with a strong band (with I$_{\rm rel.}$=0.95) due to CH stretching in the CH$_{2}$ groups can contribute to the observed 3.47 $\mu$m absorption. Even if ammonia hydrates are the predominant carriers, contributions from hydrogenated PAHs or related species cannot be excluded \citep{2023MNRAS.523.5887M}. A weak feature ($\nu_{28}$ fundamental) due to CH stretching vibrations at $\sim$3.23 $\mu$m in 3-pyrroline might contribute to the observed 3.25 $\mu$m absorption. The other strong features are for $\nu_{27}$ at 703.09 cm$^{-1}$ (14.22 $\mu$m, I$_{\rm rel.}$=0.36), $\nu_{13}$, and at 1094.56 cm$^{-1}$ (9.13 $\mu$m, I$_{\rm rel.}$=0.34).   

In the presence of water (H$_{2}$O) ices, the fundamental vibrational bands of 3-pyrroline show significant changes (Figure 7), indicating a strong interaction between the water molecules and 3-pyrroline. These interactions, likely involving hydrogen bonding and other intermolecular forces, can alter the vibrational modes of 3-pyrroline, leading to shifts in band positions and changes in band intensities. In particular, the strong OH stretching vibrations overshadow the most prominent CH stretching features of 3-pyrroline with increasing H$_{2}$O concentrations between 3000--3500 cm$^{-1}$. Furthermore, a strong fundamental band in pure 3-pyrroline at $\sim$600 cm$^{-1}$ ($\sim$16.6 $\mu$m), corresponding to NH out-of-plane bending, with other CH out-of-plane bending features between $\sim$1000--666 cm$^{-1}$ ($\sim$10--15 $\mu$m) fade away in the presence of water dimers and above, compared to the CH stretch region. The increasing concentration of H$_{2}$O highly affects the IR bands of 3-pyrroline (Figure 7), presenting challenges in drawing any definitive conclusions for discerning 3-pyrroline features in water ices. This amplifies the likelihood of more possibility of detecting it in the gas phase.

\section{Conclusions}
The recent GBT Observations of TMC1: Hunting for Aromatic Molecules (GOTHAM) survey \citep[e.g.][]{2021NatAs...5..176M,2021Sci...371.1265M} and Q-band Ultrasensitive Inspection Journey to the Obscure TMC-1 Environment
(QUIJOTE) line survey \citep[e.g.][]{2021A&A...652L...9C} suggest that an N-heterocycle similar to 3-pyrroline might exist in the ISM; however, extensive work is needed to confirm this.
The present work examines a probable formation route for 3-pyrroline in cold ISM. This work provides highly accurate rotational transitions and IR frequencies to guide laboratory and observational studies.
Pyrrolines are generally considered double-hydrogenated forms of pyrrole. However, addition of two hydrogen atoms might not be a stable endpoint. The subsequent loss of a hydrogen molecule (H$_{2}$) through abstraction could potentially lead back to the formation of pyrrole itself \citep{2021MNRAS.505.3157M}. The alternative pathway studied here suggests that 3-pyrroline can form on interstellar dust grains with successive hydrogenation of vinyl cyanide, which is the most frquently detected nitrile in space. 3-pyrroline could be searched in cold interstellar environments through its rotational transitions around 52.3 GHz. The most intense IR features of 3-pyrroline are present at 16.09 $\mu$m, and $\sim$3.5 $\mu$m, respectively. These fundamental features, located in an IR region free from PAH interference, are detectable with JWST.  
Therefore, the spectral information for the 3-pyrroline presented in this work will be crucial for both its experimental identification and future interstellar observations. Although the current work lays a strong foundation, the complexities of interstellar chemistry necessitate further investigation. 

\section*{Acknowledgements}
AP (Anshika Pandey) acknowledges Banaras Hindu University and UGC, New Delhi, India, for providing a fellowship. AV acknowledges research fellowship from DST SERB (SERB-EMR/2016/005266) and UGC. SS acknowledges RJP-PDF, Banaras Hindu University. AP (Amit Pathak) acknowledges financial support from the IoE grant of Banaras Hindu University (R/Dev/D/IoE/Incentive/2021-22/32439), financial support through the Core Research Grant of SERB, New Delhi (CRG/2021/000907) and thanks the Inter-University Centre for Astronomy and Astrophysics, Pune for associateship. KAPS acknowledges the UGC Faculty Recharge Program of Ministry of Human Resource Development (MHRD), Govt. of India and University Grants Commission (UGC), New Delhi, and support from IoE grant of Banaras Hindu University.
\section*{Data Availability}
On receiving a reasonable request from the corresponding author, the data utilized in this article will be provided.

\appendix



\bibliographystyle{mnras}
\bibliography{example} 




\bsp	
\label{lastpage}
\end{document}